\documentclass{PoS}

\def\be {\begin{equation}}
\def\ee {\end{equation}}

\title{Probing Quark Gluon Plasma by Heavy Flavors}

\ShortTitle{Probing QGP by Heavy Flavors}

\author{\speaker{Santosh K Das} and Jan-e Alam  
         \\ Theoretical Physics Division
        \\Variable Energy Cyclotron Centre, 1/AF, Bidhan Nagar, Kolkata 700064, India
        \\E-mail: \email{santosh@vecc.gov.in,jane@vecc.gov.in}}

\abstract{
The drag and diffusion coefficients of charm and bottom quarks propagating
through quark gluon plasma (QGP) have been evaluated within the framework of perturbative
Quantum Chromodynamics (pQCD). Both radiative and collisional processes of dissipation are included
in evaluating these transport coefficients. The dead cone as well as the LPM effects on 
radiative energy loss of heavy quarks have also been  considered. The Fokker Planck equation  
has been solved to study the dissipation of heavy quarks momentum in QGP.
The nuclear suppression factor, $R_{\mathrm AA}$  and the elliptic flow $v_2^{HF}$ of the
non-photonic electrons resulting from the semi-leptonic decays 
of hadrons containing charm and bottom quarks have been 
evaluated  for  RHIC and  LHC nuclear 
collision conditions. We find that the observed $R_{\mathrm AA}$  and $v_2$ at RHIC  
can be reproduced simultaneously within the pQCD framework.
}
\FullConference{The Seventh Workshop on Particle Correlations and Femtoscopy\\
		 September 20 - 24 2011\\
		 University of Tokyo, Japan}

\begin{document}
\section{Introduction}

The nuclear collisions at Relativistic Heavy Ion Collider (RHIC) and
Large Hadron Collider (LHC) energies are aimed at creating a
new state of matter - properties of which are governed by 
thermal quarks and gluons, such a phase of matter
is called Quark-Gluon Plasma(QGP). The study of
the properties of QGP is a field of high contemporary interest and the
heavy quarks (HQ) namely, charm and bottom quarks play crucial roles
in this endeavour. Because HQs (i) are produced in the early stage of the
collisions and hence witness the entire evolution scenario of the matter,
(ii) do not dictate the bulk properties of the matter (iii) their thermalization time 
is larger than light quarks and gluons and 
therefore, the propagation of HQs through a QGP may be
treated as the interactions between equilibrium and non-equilibrium
degrees of freedom. The Fokker-Planck equation (FPE) provides an appropriate
framework for such studies. In the present work we evaluate 
the nuclear suppression factor and elliptic flow of 
heavy flavours by solving the FPE for the motion of 
charm and bottom quarks in QGP and contrast the theoretical results 
with the data to extract the properties of QGP.

\section{Formalism}
The evolution of HQs momentum distribution function while propagating
through the QGP can be described by the FPE, 
which reads~\cite{FP},

\begin{eqnarray}
\frac{\partial f}{\partial t} = 
\frac{\partial}{\partial p_i} \left[ A_i(p)f + 
\frac{\partial}{\partial p_j} \lbrack B_{ij}(p) f \rbrack\right] 
\label{expeq}
\end{eqnarray}
where the kernels $A_i$ and $B_{ij}$ are given by
\begin{eqnarray}
&& A_i = \int d^3 k \omega (p,k) k_i \nonumber\\
&&B_{ij} = \int d^3 k \omega (p,k) k_ik_j.
\end{eqnarray}
The function $\omega(p,k)$ 
is given by
\begin{eqnarray}
\omega(p,k)=g_j\int\frac{d^3q}{(2\pi)^3}f_j(q)v_{ij}\sigma^j_{p,q\rightarrow p-k,q+k}
\end{eqnarray}
where $g_j$ is the statistical degeneracy, $f_j$ is the phase space distribution of the particle $j$,
$v_{ij}$ is the relative velocity between the two collision partners and
$\sigma$ denotes the cross section.  
For $\mid\bf{p}\mid\rightarrow 0$,  $A_i\rightarrow \gamma p_i$
and $B_{ij}\rightarrow D\delta_{ij}$ where $\gamma$ and $D$ stand for
drag and diffusion co-efficients respectively. 

As mentioned before we include both the collisional and 
radiative processes for HQs dissipation in QGP. 
 For the collisional processes~\cite{Das1,Das2}: 
$gQ \rightarrow gQ$ and $qQ \rightarrow qQ$ and $\bar{q}Q \rightarrow \bar{q}Q$ are 
considered in evaluating the drag ($\gamma_{coll}$) and diffusion ($D_{coll}$) coefficients. 

For the radiative loss the drag coefficient is defined  through the relation,
\begin{eqnarray}
\frac{-dE}{dx}=\gamma_{rad} p
\end{eqnarray}
and use the Einstein relation $D_{rad} = TM\gamma_{rad}$  to obtain the diffusion coefficient,
where  $M$ is the HQ mass,  $T$ is the temperature of the QGP. 
In the radiative process the dead cone
and Landau-Pomerenchuk-Migdal (LPM) effects are included~\cite{Das1}.
Although the collisional and radiative processes are not independent from each other, 
however, in the absence of any rigorous method, we add them up to obtain 
the effective drag coefficients, $\gamma_{eff} = \gamma_{rad} + \gamma_{coll}$, 
and similarly the effective diffusion coefficient, $D_{eff} = D_{rad} + D_{coll}$. 
This is a good approximation for the present work because the radiative 
loss is large compared to the collisional loss. 
In evaluating the drag co-efficient we have
used temperature dependent  strong coupling,
$\alpha_s$~\cite{zantow}. The Debye mass, $\sim g(T)T$ is  also a temperature dependent
quantity used here to shield the infrared divergences
arising due to the exchange of massless gluons.
The FPE has been solved with these effective 
transport coefficients and the initial momentum distributions of 
HQs are taken from the proton-proton (pp) collisions~\cite{MNR}  
at RHIC and LHC energies.

The solution of the FPE are convoluted with the fragmentation functions~\cite{peterson}
of the HQs to obtain the $p_T$ distribution of the 
$D$ and $B$ mesons. 
The transverse momentum spectra of the electrons originating from the
decays:  $D\rightarrow X e \nu$ and $B\rightarrow X e \nu$  have
been obtained by using standard procedures~\cite{cls}.

(2+1) dimensional relativistic hydrodynamics~\cite{hydro}
with boost invariance~\cite{bjorken} along
the longitudinal direction has been used for the 
space time description of the of the flowing QGP. We have taken the 
Equation of State (EoS) as $P=c_s^2\epsilon$, where $P$ is the 
pressure, $\epsilon$ is the energy density and $c_s$ is the velocity of sound in QGP.
Variation of $R_{AA}$ with $c_s^2$ {\it i.e.} with EoS will be presented
here. We have assumed the value of the
initial temperature, $T_i=400$ MeV and thermalization time of the QGP, 
$\tau_i=0.2$ fm/c for Au+Au collisions at RHIC. 
The corresponding values for $T_i$ and $\tau_i$ for LHC are taken as
700 MeV and 0.08 fm/c respectively. 
It is expected that the central rapidity
region of the system formed in nuclear collisions
at RHIC and LHC energies is almost net baryon density free. Therefore,
the equation governing the conservation of net
baryon number need not be considered here.

The total amount of energy dissipated by a  HQ in the QGP
depends on the path length it traverses.
Each parton traverse different path length
which depends on the  geometry of the system and on the point 
where its is created.
The probability that a parton is produced at a point $(r,\phi^\prime)$
in a QGP of ellipsoidal shape depends on the number of binary collisions 
at that point which can be taken as:
\be
P(r,\phi^\prime)=\frac{1}{{\cal N}}\left(1-\frac{r^2}{R^2}
\frac{(1+\epsilon cos^2\phi^\prime)}{(1-\epsilon^2)^2}\right)\Theta(R-r)
\label{eq7}
\ee
and
\be
{\cal{N}}=\frac{1}{\pi R^2\left(1-\frac{1}{2}\frac{1+\epsilon^2/2}
{(1-\epsilon^2)^2}\right)
}
\label{eq8}
\ee
where $R$ is the nuclear radius 
and $\epsilon$ is the (spatial) eccentricity of the ellipse. 
A parton created at $(r,\phi^\prime)$ in the transverse plane
propagate a distance $L=\sqrt{R^2-r^2sin^2\phi^\prime}-rcos\phi^\prime$
in the medium. We use the following
equation for the geometric average of the integral which appear
in the solution of the FPE~\cite{Das2}
involving drag coefficient: 
\be
\Gamma=\frac{\int rdr d\phi^\prime P(r,\phi^\prime) \int^{L/v}d\tau\gamma(\tau)}
{\int rdr d\phi^\prime P(r,\phi^\prime)}
\label{eq9}
\ee
where $v$ is the velocity of the propagating partons. 
Similar averaging has been performed  for the diffusion co-efficient.

\section{Results}
In this section we contrast the theoretical results  with the experimental
data available for Au+Au collisions at RHIC energies and also present 
theoretical results for LHC energy.

\subsection{The Nuclear Suppression}
\begin{figure}[ht]
\begin{center}
\includegraphics[scale=0.35]{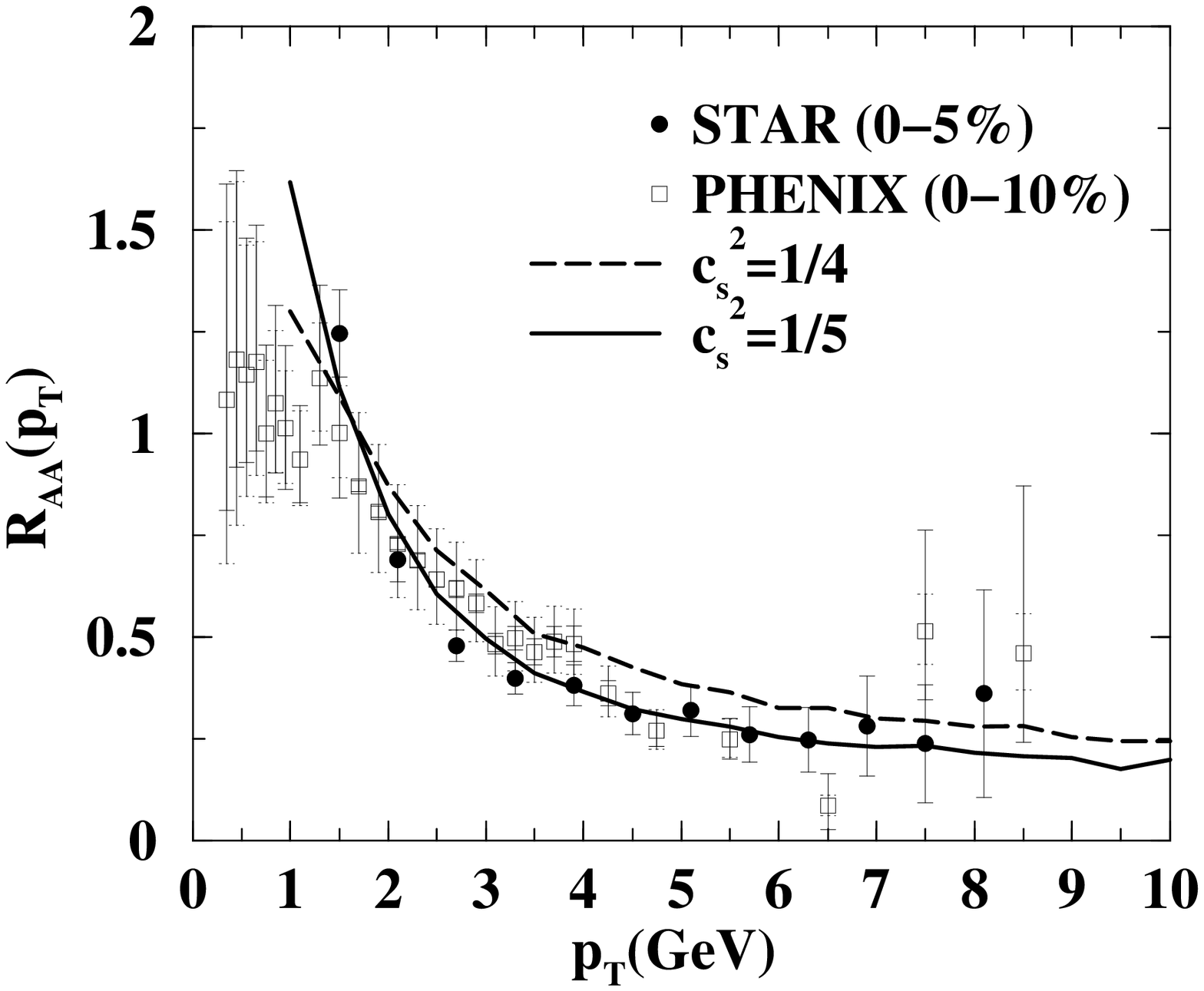}
\includegraphics[scale=0.35]{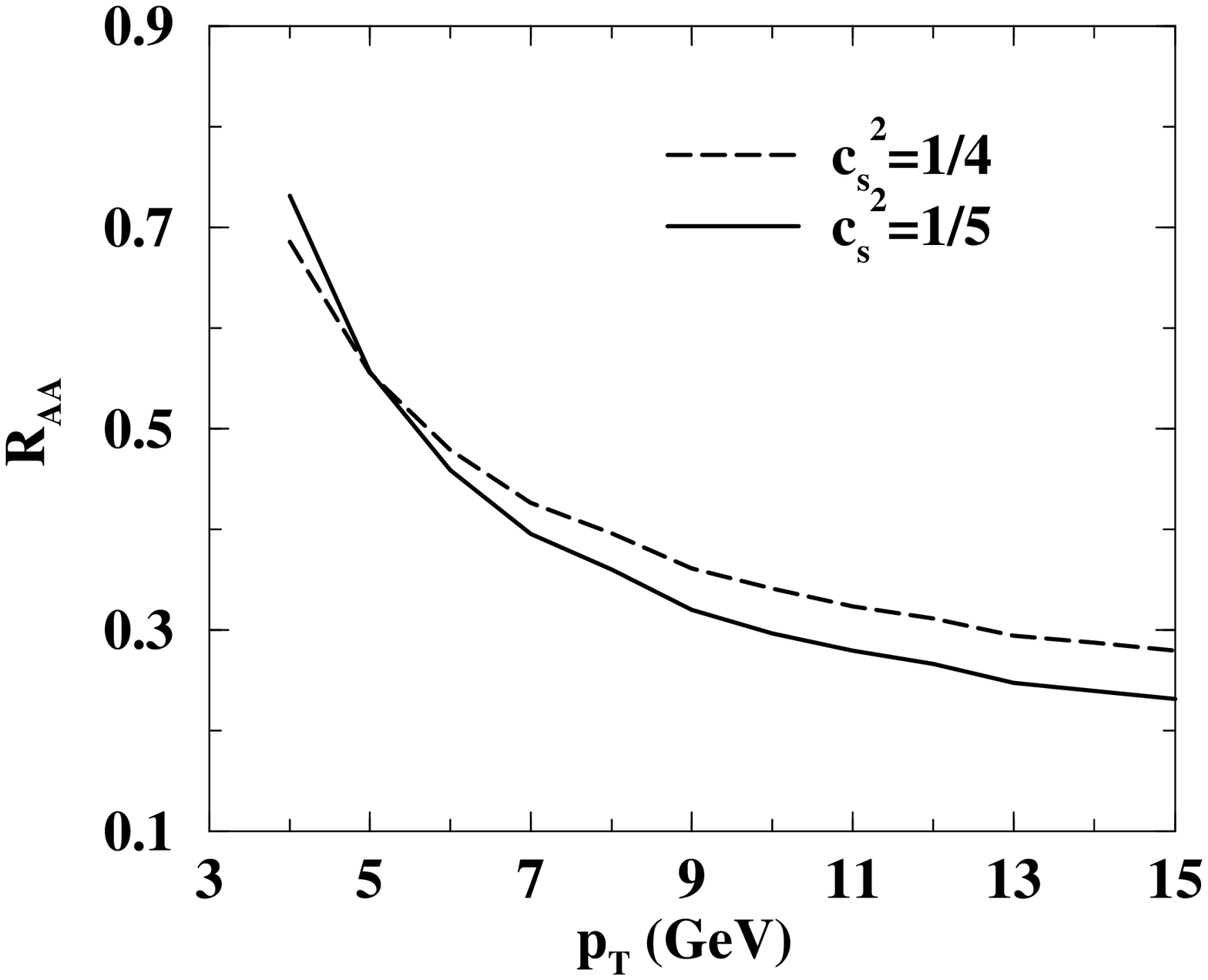}
\caption{Left panel: variation of $R_{\mathrm AA}$ with $p_T$ obtained from the
present work compared 
with the experimental data measured by STAR and PHENIX collaboration
at $\sqrt{s_{\mathrm NN}}=200$ GeV. The experimental data of STAR
and PHENIX collaborations are
taken from ~\cite{stare} and~\cite{phenixe} respectively. 
Right panel: variation of $R_{AA}$ with $p_T$ for LHC.
 }
\label{fig1}
\end{center}
\end{figure}

The transverse momentum distribution of electrons from the heavy flavours produced in
p-p collisions can be estimated from the charm and beauty quarks
distributions which provides initial condition to the FPE.
The solution of FPE for the HQ transverse momentum distribution contains 
the effects of drag (quenching)
on the HQ whereas the initial distributions of HQ
does not contain any such effects, therefore the ratio
of these two quantities, the nuclear suppression factor,
$R_{AA}$ acts as a marker for the momentum degradation in the medium, which
is observed experimentally through, 
$R_{AA}(p_T)$ defined as:
\begin{eqnarray}
R_{AA}(p_T)=\frac{\frac{dN^e}{d^2p_Tdy}^{\mathrm Au+Au}}
{N_{\mathrm coll}\times\frac{dN^e}{d^2p_Tdy}^{\mathrm p+p}}
\label{raa}
\end{eqnarray}
The experimental data from both the PHENIX and STAR collaborations~\cite{stare,phenixe}
shows substantial suppression ($R_{AA}<1$) for $p_T\geq 2$ GeV indicating
significant dissipation of HQ energy in the QGP. 
The theoretical results describe the data reasonably (Fig.~\ref{fig1}, left panel). In
the right panel we display the results for LHC for two values of $c_s$.  
We have taken an equation of state
with velocity of sound lower than
the value corresponding to the Stefan-Boltzmann limit.
Lower value of $c_s$ makes the expansion of the plasma slower
enabling the propagating heavy quarks to spend more time to interact
in the medium and hence lose more energy before exiting from the plasma.
\subsection{Elliptic Flow}
The elliptic flow, $v_2^{HF}$ is defined as:
\begin{eqnarray}
v_2^{HF}(p_T)=\langle cos(2\phi) \rangle= \frac{\int d\phi \frac{dN}{dydp_Td\phi×}|_{y=0} cos(2\phi)}
{\int d\phi \frac{dN}{dydp_Td\phi×}|_{y=0}×}
\end{eqnarray}

\begin{figure}[ht]
\begin{center}
\includegraphics[scale=0.35]{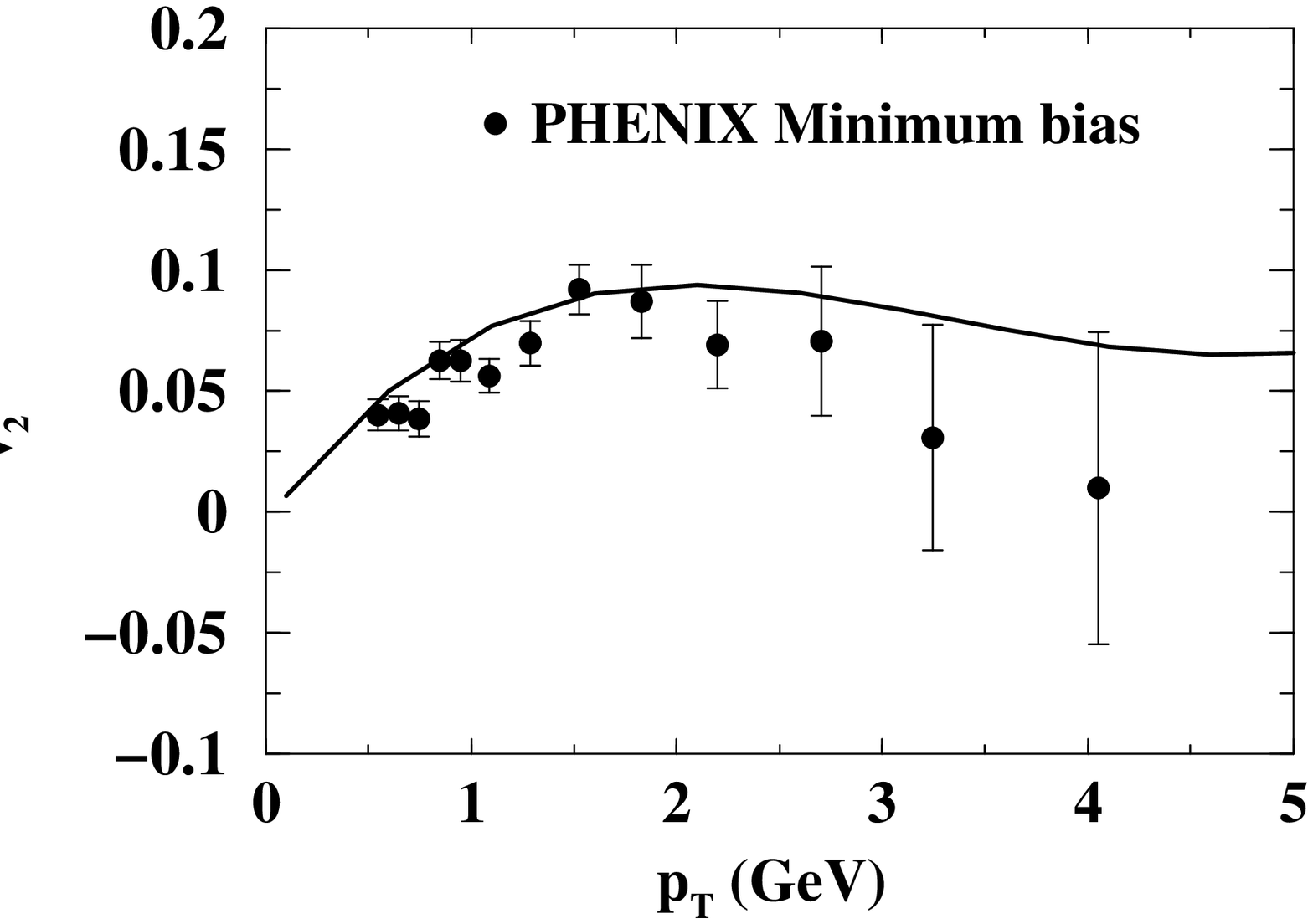}
\includegraphics[scale=0.35]{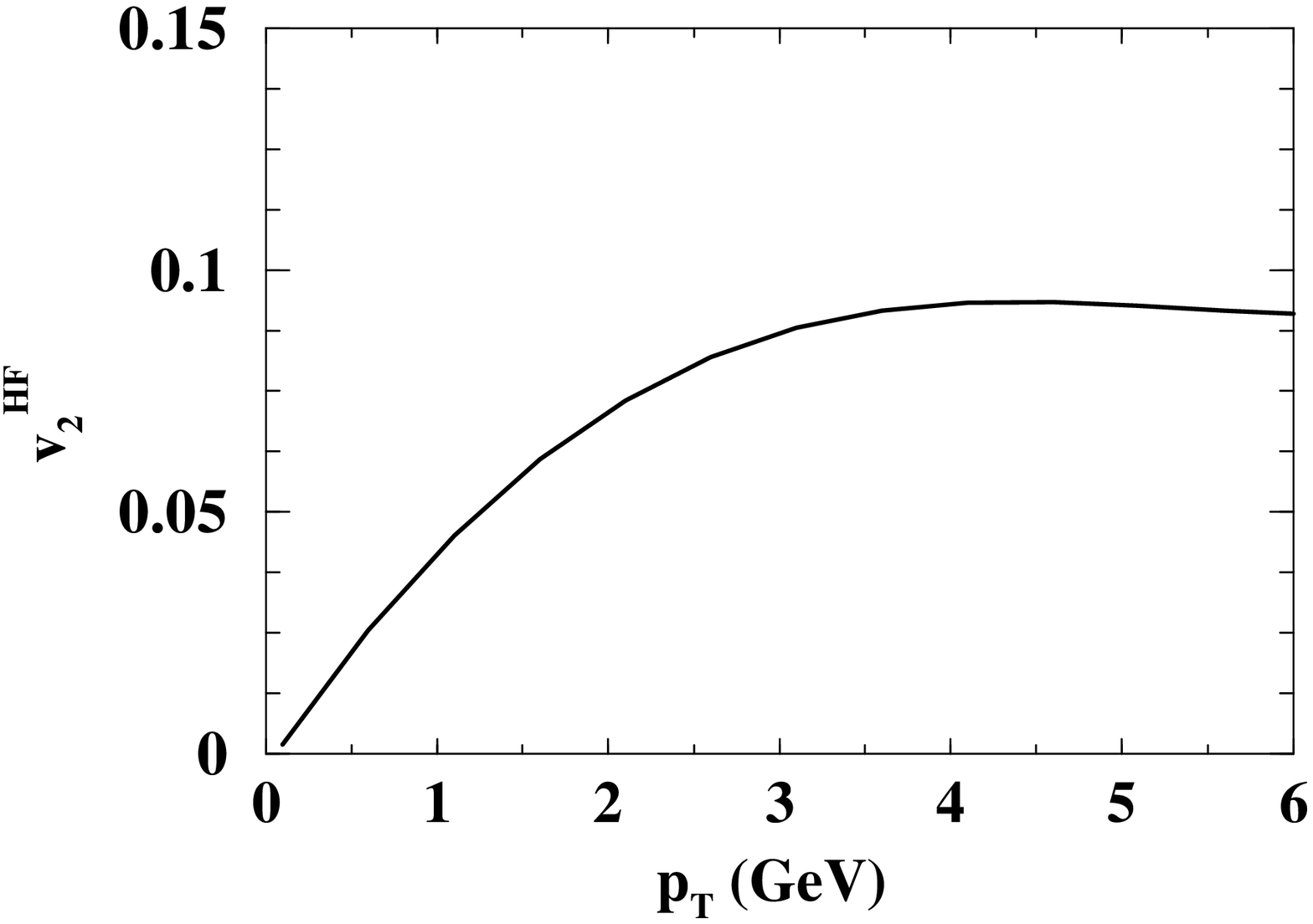}
\caption{Left panel: variation of $v_2^{HF}$ with $p_T$ at  the
highest RHIC energy. Experimental data is taken from~\cite{phenixe}.
The value of the ``effective'' impact parameter, $b=10.2$ fm/c for 
minimum bias Au+Au collision at $\sqrt{s_{\mathrm NN}}=200$ GeV.
Right panel: variation of $v_2^{HF}$ with $p_T$ for LHC. 
}
\label{fig2}
\end{center}
\end{figure}

In the left panel of Fig.~\ref{fig2} we contrast the experimental data obtained by  the
PHENIX~\cite{phenixe} collaborations 
for Au + Au minimum bias collisions at $\sqrt{s_{\mathrm NN}}=200$ GeV with
theoretical results obtain in the present work. 
It is observed that the $v_2^{HF}$ first increases and reaches a maximum of about 8\% 
then  saturates for  $p_T>2$ GeV.
We also find that the 
data can be reproduced well by including both radiative and collisional loss
with $c_s=1/\sqrt{4}$.  In the right panel we display the variation of $v_2^{HF}$ with $p_T$
for LHC energies with $c_s=1/\sqrt{4}$.  

\section{Summary and discussions}
In summary, we have simultaneously reproduced  
the measured  nuclear suppression and elliptic flow
of heavy flavours at RHIC energies within the framework
of pQCD interactions of the non-equilibrated HQ with flowing QGP. 
We observe that the inclusion of the collisional and radiative processes
of dissipation and use of non-ideal EoS 
are the  key factors for the successful reproduction of the data.
Our analysis admits the formation of QGP in Au+Au collisions at
centre of mass energy, $\sqrt{s_{\mathrm NN}}=200$ GeV with initial temperature
$\sim 400$ MeV. 

Some comments on the $R_{\mathrm AA}$  vis-a-vis $v_2^{HF}$
are in order here.  The 
$R_{\mathrm AA}$  contains the ratio of $p_T$ distribution 
of the electrons resulting from Au+Au to p+p collisions, while
the numerator contains the interaction of the HQs
with the flowing QGP,  such interactions are absent in 
the denominator. Whereas for $v_2^{HF}$ both the numerator and the
denominator contain the interactions with the medium, resulting in
some sort of cancellation. 
Therefore, $K$ factor which is sometimes 
used for the reproduction of the data may not be 
same for $R_{\mathrm AA}$ and $v_2^{HF}$. 

Several theoretical attempts have been made to
explain $R_{\mathrm AA}$ and $v_2$, where the role of 
hadronic matter has been ignored. However, to make the
characterization of QGP reliable the role of the
hadronic phase should be taken into consideration 
and its contribution must be subtracted out from the observables.
Recently the drag and 
diffusion coefficients of hot hadronic medium consisting of
pions, kaons and eta using open charm and beauty mesons 
as a probe have been evaluated~\cite{Ghosh, Das3,laine,MinHe,abreu}. 
It is observed that the magnitude of both the transport coefficients  
are significant, indicating substantial interactions of the
heavy mesons with hot hadrons, which may have significant
consequences of $R_{AA}$ and $v_2^{HF}$.

{\bf Acknowledgment:} 
SKD is supported by DAE-BRNS project Sanction No.  2005/21/5-BRNS/2455.

\end{document}